\documentstyle[prb,aps,epsf]{revtex}

\begin{document}
\draft


\title{Vortex Charging Effect in a Chiral $p_x\pm i p_y$-Wave Superconductor}

\author{Masashige Matsumoto$^1$ and Rolf Heeb$^2$}
\address{$^1$Department of Physics, Faculty of Science, Shizuoka University, 836 Oya, Shizuoka 422-8529, Japan \\
$^2$Theoretische Physik, Eidgen\"ossische Technische Hochschule H\"onggerberg, CH-8093 Z\"urich, Switzerland}

\date{\today}
\maketitle

\begin{abstract}
Quasiparticle states around a single vortex in a $p_x\pm i p_y$-wave superconductor
are studied on the basis of the Bogoliubov-de Gennes (BdG) theory,
where both charge and current screenings are taken into account.
Due to the violation of time reversal symmetry,
there are two types of vortices which are distinguished by their winding orientations
relative to the angular momentum of the chiral Cooper pair.
The BdG solution shows that the charges of the two types of vortices are quite different,
reflecting the rotating Cooper pair of the $p_x\pm i p_y$-wave paring state.
\end{abstract}


\vspace*{10pt}


\section{Introduction}

The discovery of many types of superconductors from heavy fermion compounds to high-$T_c$ cuprates
has driven us to study a large variety of new physics
beyond the standard BCS theory for conventional $s$-wave superconductors.
The study of the unconventional superconductivity was stimulated
by the discovery of superfluid $^3$He in which a spin triplet $p$-wave state realizes.
Unlike the conventional $s$-wave state,
the $p$-wave state has both spin and orbital degrees of freedom.
This is the most pronounced feature of the unconventional superconductors
observed in their thermodynamics and impurity effects
or detected by tunneling spectroscopy, NMR, $\mu$SR measurements, and so on.

Sr$_2$RuO$_4$ is the first layered perovskite compound
showing superconductivity without CuO$_2$ planes.
\cite{Maeno94}
The recent experimental and theoretical studies have indicated
that the superconducting pairing symmetry of Sr$_2$RuO$_4$ is not a simple $s$-wave.
The absence of a Hebel-Slichter peak in NQR
\cite{Ishida97}
and the sensitivity of $T_c$ on non-magnetic impurities
\cite{Mackenzie98}
point towards unconventional pairing. 
The indication of broken time reversal symmetry,
\cite{Luke98}
observed in $\mu$SR measurements,
gives a strong argument for the unconventional paring state.
The Knight shift experiment shows that the spin susceptibility
is not affected by the superconducting state,
\cite{Ishida98}
which is the strong evidence of a spin-triplet pairing.
Sigrist {\it et al}. suggested that a $p_x\pm ip_y $-wave state,
which breaks the time reversal symmetry in a tetragonal crystal field,
is the most likely pairing state for Sr$_2$RuO$_4$.
\cite{Sigrist99}
The line node behavior is reported in the latest experiments
\cite{Yoshida99,NishiZaki00,Ishida00}
related to the low temperature thermodynamical measurements,
such as specific heat and NMR $T_1^{-1}$.
An orbital dependent superconductivity \cite{Agterberg97} and gap anisotropy
\cite{Miyake99,Hasegawa00}
were suggested to understand the line node behavior.
Here we focus on the $p_x\pm i p_y$-wave pairing state,
since this representation is the simplest and essential form.
We will see rich physics of this chiral state.

The most intriguing character of the $p_x\pm i p_y$-wave state is
that the Cooer pair has $\pm 1$ angular momentum,
i.e. the pair electrons are rotating.
This property is similar to that of the A phase of the superfluid $^3$He.
Due to the violation of the time reversal symmetry, we have two types of vortices,
one of which is in the same direction to the angular momentum of the rotating Cooper pair,
and the other is in the opposite direction.
The rotating pair shows up in quasiparticle states around a vortex core.

In this letter we present an interesting new physics related to a vortex.
We focus on a vortex charging effect.
\cite{Khomskii95}
It was pointed out that for an $s$-wave superconductor the vortex charge is
proportional to the slope of the density of states at the Fermi level.
\cite{Blatter96}
Hayashi {\it et al.} has proposed that the vortex charge is determined
by the quasiparticle structure in the vortex core
\cite{Hayashi98-2}
rather than the slope of the density of states.
Very recently, it has been reported that Chern-Simons terms lead to a fractional vortex charge
for the $p_x\pm i p_y$-wave state.
\cite{Goryo99}
Thus the origin of the vortex charge is still controversial.
To make this point clearer,
we investigate the vortex charging effect in the chiral $p_x\pm i p_y$-wave state,
concentrating our attention on a microscopic origin of the vortex charge.
This is demonstrated by solving the BdG equation self-consistently,
including both charge and current screenings.
This is the first full self-consistent BdG study of a single vortex.
We would like to show how the rotating Cooper pair shows up in the vortex charging effect.

This paper is organized as follows.
In Sec. II, we present our formulation
for calculating the quasiparticle states near the vortex core based on the BdG equation.
In Sec. III, we show numerical results.
We give summary and discussions in Sec. IV.

\section{Formulation}
Let us begin with the following BdG and gap equations
for the $p_x\pm i p_y$-wave state:
\cite{Matsumoto99,Heeb-doctor,Furusaki}
\begin{mathletters}
\begin{eqnarray}
&&h_0 u_n - \frac{i}{k_F}
  \sum_\pm \left[ \Delta_\pm \Box_\pm + \frac{1}{2} (\Box_\pm \Delta_\pm) \right] v_n
  =E_n u_n, \\
-&&h_0^* v_n - \frac{i}{k_F}
  \sum_\pm \left[ \Delta_\pm \Box_\pm + \frac{1}{2} (\Box_\pm \Delta_\pm) \right]^* u_n
  = E_n v_n, \\
&&h_0 = \frac{1}{2m}\left(-i \nabla + \frac{e}{c}\bbox{A}\right)^2 -eA_0-\mu, \\
&&\Delta_\pm (\bbox{r})
  = - i \frac{V_p \Omega}{2 k_F}
  \sum_{0 \le E_n \le E_{\rm c}} \tanh\left(\frac{E_n}{2T}\right)
  \left[ v_n^*(\bbox{r}) \Box_\mp u_n (\bbox{r})
       - u_n(\bbox{r}) \Box_\mp v_n^* (\bbox{r}) \right].
\end{eqnarray}
\label{eqn:BdG}
\end{mathletters}\noindent
The detailed derivation is given in the Appendix.
Throughout this paper, we use $\hbar=1$ and $k_{\rm B}=1$ units.
$e>0$ is the electron charge.
$E_n$ is the energy eigenvalue for the superconducting quasiparticle,
and $\Box_\pm = \partial_x \pm i \partial_y$ is used.
$A_0$ and $\bbox{A}$ are the scalar and vector potentials, respectively.
$V_p>0$ and $\Omega$ are the $p$-wave attractive interaction
and volume of the two-dimensional system, respectively.
$E_{\rm c}$ is a cutoff energy, and $\mu$ is a chemical potential.
$\Delta_\pm$ is the order parameter for the $p_x\pm i p_y$-wave.
For simplicity, we assume that the superconductor is basically two-dimensional
and has a cylindrical Fermi surface.
We note that the BdG solution has the following time reversal relation:
\begin{equation}
\{u_{-E_n},v_{-E_n}\} = \{v_{E_n}^*,u_{E_n}^*\}.
\label{eqn:symmetry}
\end{equation}
The solution of the BdG equation determines the two-dimensional electron and current densities:
\begin{mathletters}
\begin{eqnarray}
n(\bbox{r}) &=&
  2 \sum_{E_n} |u_n(\bbox{r})|^2 f(E_n), \\
\bbox{J}(\bbox{r}) &=&
  -\frac{ie}{m} \sum_{E_n} \left[ u_n^*(\bbox{r})\nabla u_n(\bbox{r})
                            -u_n(\bbox{r})\nabla u_n^*(\bbox{r}) \right] f(E_n)
  -\frac{e^2}{mc} n(\bbox{r})\bbox{A}(\bbox{r}).
\end{eqnarray}
\label{eqn:density}
\end{mathletters}\noindent
By using the relation (\ref{eqn:symmetry}),
both densities have been expressed in terms of the $u_n$ amplitude.
$f(E_n)$ is the Fermi distribution function
and the $E_n$ summation in Eq. (\ref{eqn:density}) runs both negative and positive regions.
The scalar and vector potentials obey the Maxwell equations:
\begin{mathletters}
\begin{eqnarray}
\nabla^2 A_0(\bbox{r}) &=& -\frac{4\pi e}{d} \left[ n_0-n(\bbox{r})\right], \\
\nabla^2 \bbox{A}(\bbox{r}) &=& -\frac{4\pi}{cd} \bbox{J}(\bbox{r}).
\end{eqnarray}
\end{mathletters}\noindent
We have introduced a layer spacing $d$ to convert the area densities into the volume ones.
To satisfy charge neutrality we introduce a uniform $n_0$
as the density of positive background charge.
We have taken the origin of the coordinate at the vortex center.

In the bulk region $p_x +i p_y$ and $p_x -i p_y$-wave states are degenerate.
In this letter we will choose one of the two degenerate states, the $p_x +i p_y$-wave state
as a dominant component.
However, the other component ($p_x -i p_y$) is admixed with the bulk state ($p_x +i p_y$)
close to the vortex core.
\cite{Agterberg98}
Therefore our formulation includes both $p_x \pm i p_y$ components.
The $p_x + i p_y$-wave state has a $+1$ Cooper pair phase winding,
since $\Box_+$ in Eq. (\ref{eqn:BdG}) is expressed as
\begin{equation}
\Box_+ = e^{i \phi} (\partial_r + \frac{i}{r}\partial_\phi).
\end{equation}
Here $r$ and $\phi$ are the two-dimensional polar coordinates.
A single vortex produces an additional phase winding $\pm 1$.
Therefore there are two kind of vortices:
the vortex winding direction is parallel or anti-parallel relative to the Cooper pair winding.
As we will see later, the former vortex is charged up, while the latter is not.
We name the former and the latter C(charged)-vortex and U(uncharged)-vortex, respectively.
The C-vortex has two fold winding $+2$ effectively.
On the other hand,
the phase winding is zero for the U-vortex,
since the winding of the Cooper pair cancels the vortex one.
Due to the rotational symmetry of the system,
angular momentum is a good quantum number.
The BdG solutions are then classified by the angular momentum.
To keep the rotational symmetry,
there is only the following combination of the order parameters for the C and U-vortices:
\begin{mathletters}
\begin{eqnarray}
({\rm C-vortex})~~~~~~&&\Delta_+(r,\phi) = \Delta_+(r) e^{i \phi},
~~~~~~\Delta_-(r,\phi) = \Delta_-(r) e^{i 3\phi}, \\
({\rm U-vortex})~~~~~~&&\Delta_+(r,\phi) = \Delta_+(r) e^{-i \phi},
~~~~~~\Delta_-(r,\phi) = \Delta_-(r) e^{i \phi},
\end{eqnarray}
\end{mathletters}\noindent
where $\Delta_+$ and $\Delta_-$ are the dominant and admixed components, respectively.
Hence the wavefunction $u_n$ couples with $v_n$ in the following angular momentum spaces:
\cite{Matsumoto99}
\begin{eqnarray}
&&~~~({\rm C-vortex}) ~~~u_n^l \leftrightarrow v_n^{l-2}, \\
&&~~~({\rm U-vortex}) ~~~u_n^l \leftrightarrow v_n^l,
\end{eqnarray}
where the superscripts represent the angular momenta of $u_n$ and $v_n$.
The GL calculation showed that the U-vortex is energetically favored.
\cite{Heeb99}
However, in a real sample we can expect two types ($p_x\pm i p_y$) of domains,
so that there are both C and U-vortices in the presence of an external magnetic field.
We solve the two-dimensional single vortex problem on a disk of radius $R$.
Since we treat a cylindrical system,
it is convenient to expand wave functions by the following base functions:
\begin{mathletters}
\begin{eqnarray}
\left(
  \matrix{
    u_n(\bbox{r}) \cr
    v_n(\bbox{r}) \cr
  }
\right)
&=&
\sum_{j}
\left(
  \matrix{
    u_{lnj} e^{i l\phi}\varphi_{lj}(r) \cr
    v_{l'nj} e^{i l'\phi}\varphi_{l'j}(r) \cr
  }
\right), \\
\varphi_{lj}(r) &=& \frac{1}{\sqrt{\pi R^2}} J_l({Z_{lj}r \over R})
\end{eqnarray}
\label{eqn:Bessel}
\end{mathletters}\noindent
Here $J_l$ is the $l$-th Bessel function,
$Z_{lj}$ is the $j$-th zero of $J_l$
and $R$ is the radius of the system.
$l'$ in Eq. (\ref{eqn:Bessel}a) takes $l-2$ ($l$) for the C-vortex (U-vortex).

In the practical numerical calculation, we use no-dimensional quantities.
We express them with bar as follows:
\begin{eqnarray}
\nabla &=& \frac{1}{\xi_0} \bar{\nabla},~~~~~~T = \Delta_0 \bar{T}, \cr
\Delta_\pm &=& \Delta_0 \bar{\Delta}_\pm,~~~~~~E_n = \Delta_0 \bar{E}_n, \cr
\mu &=& \Delta_0 \bar{\mu},~~~~~~E_{\rm c} = \Delta_0 \bar{E}_c, \cr
\bbox{r} &=& \xi_0 \bar{\bbox{r}},~~~~~~\Box_\pm = \frac{1}{\xi_0} \bar{\bar{\Box}}_\pm, \cr
u_n &=& \frac{1}{\xi_0} \bar{u}_n,~~~~~~v_n = \frac{1}{\xi_0} \bar{v}_n, \cr
e A_0 &=& \Delta_0 \bar{a}_0,~~~~~~\frac{e}{c} \bbox{A} = \frac{1}{\xi_0} \bar{\bbox{a}}, \cr
n &=& \frac{1}{\xi_0^2} \bar{n},~~~~~~n_0 = \frac{1}{\xi_0^2} \bar{n}_0, \cr
\bbox{J} &=& \frac{e}{m \xi_0^3} \bar{\bbox{J}}.
\nonumber
\end{eqnarray}
Here $\Delta_0$ (unit of energy) is the magnitude of the order parameter at $T=0$.
$\xi_0=v_{\rm F}/\Delta_0$ (unit of length) is the superconducting coherence length.
Using these non-dimensional quantities, we obtain the following set of equations:
\begin{mathletters}
\begin{eqnarray}
\bar{h}_0 \bar{u}_n &-& \frac{i}{k_F \xi_0}
  \sum_\pm \left[ \bar{\Delta}_\pm \bar{\bar{\Box}}_\pm
                 + \frac{1}{2} (\bar{\bar{\Box}}_\pm \bar{\Delta}_\pm) \right] \bar{v}_n
  = \bar{E}_n \bar{u}_n, \\
-\bar{h}_0^* \bar{v}_n &-& \frac{i}{k_F \xi_0}
  \sum_\pm \left[ \bar{\Delta}_\pm \bar{\bar{\Box}}_\pm
                 + \frac{1}{2} (\bar{\bar{\Box}}_\pm \bar{\Delta}_\pm) \right]^* \bar{u}_n
  = \bar{E}_n \bar{v}_n, \\
\bar{h}_0 &=& \frac{1}{2 k_{\rm F} \xi_0}\left(-i \bar{\nabla} + \bar{\bbox{a}}\right)^2
           -\bar{a}_0-\bar{\mu}, \\
\bar{\Delta}_\pm (\bar{\bbox{r}})
  &=& - i g \frac{1}{2 k_F \xi_0}
  \sum_{0 \le \bar{E}_n \le \bar{E}_{\rm c}} \tanh\left(\frac{\bar{E}_n}{2\bar{T}}\right)
        \left[ \bar{v}_n^*(\bar{\bbox{r}}) \bar{\bar{\Box}}_\mp \bar{u}_n (\bar{\bbox{r}})
       - \bar{u}_n(\bar{\bbox{r}}) \bar{\bar{\Box}}_\mp \bar{v}_n^* (\bar{\bbox{r}}) \right], \\
\bar{n}(\bar{\bbox{r}}) &=&
  2 \sum_{\bar{E}_n} |\bar{u}_n(\bar{\bbox{r}})|^2 f(\bar{E}_n), \\
\bar{\bbox{J}}(\bar{\bbox{r}}) &=&
  -i \sum_{\bar{E}_n} \left[ \bar{u}_n^*(\bar{\bbox{r}})\bar{\nabla} \bar{u}_n(\bar{\bbox{r}})
        -\bar{u}_n(\bar{\bbox{r}})\bar{\nabla} \bar{u}_n^*(\bar{\bbox{r}}) \right] f(\bar{E}_n)
        - \bar{n}(\bar{\bbox{r}})\bar{\bbox{A}}(\bar{\bbox{r}}), \\
\bar{\nabla}^2 \bar{a}_0(\bar{\bbox{r}}) &=& -\frac{\pi}{k_{\rm F}\xi_0}
             \left(\frac{\xi_0}{\lambda_{\rm TF}}\right)^2
             \left[ \bar{n}_0-\bar{n}(\bar{\bbox{r}})\right], \\
\bar{\nabla}^2 \bar{\bbox{a}}(\bar{\bbox{r}}) &=& -\left(\frac{\xi_0}{\lambda_{\rm L}}\right)^2
      \frac{1}{\bar{n}_0} \bar{\bbox{J}}(\bar{\bbox{r}}).
\end{eqnarray}
\label{eqn:BdG2}
\end{mathletters}\noindent
Here $\lambda_{\rm TF}$ and $\lambda_{\rm L}$ are the Thomas-Fermi screening lenth
and London penetration depth, respectively.
They are given by
\begin{mathletters}
\begin{eqnarray}
\lambda_{\rm TF}^2 &=& \frac{1}{4\pi e^2 N(0)} = \frac{d}{4 e^2 m}, \\
\lambda_{\rm L}^2 &=& \frac{c^2}{4\pi} \frac{m d}{e^2 n_0},
\end{eqnarray}
\end{mathletters}\noindent
where $N(0)$ is the density of states at Fermi energy.
For two-dimensional layer system, it is given by $N(0)=m/\pi d$.
$g$ in Eq. (\ref{eqn:BdG2}d) is a non-dimensional coupling for the $p$-wave defined as
$g=V_p \Omega/(\Delta_0 \xi_0^2)$.
We have three no-dimensional parameters:
$k_{\rm F}\xi_0$, $\lambda_{\rm TF}/\xi_0$, and $\lambda_{\rm L}/\xi_0$.

In our numerical calculation, we obtain $\bar{\Delta}_\pm(\bar{\bbox{r}})$,
$\bar{a}_0(\bar{\bbox{r}})$, and $\bar{\bbox{a}}(\bar{\bbox{r}})$
by solving Eq. (\ref{eqn:BdG2}) iteratively within a Gygi and Schl\"uter method.
\cite{Gygi91}
In the self-consistent calculation we fix the total number of electrons to the normal state value
by adjusting $\bar{\mu}$.

\section{Numerical Results}
\begin{figure}[t]
\begin{center}
\begin{minipage}{7cm}
\epsfxsize=7cm
\epsfbox{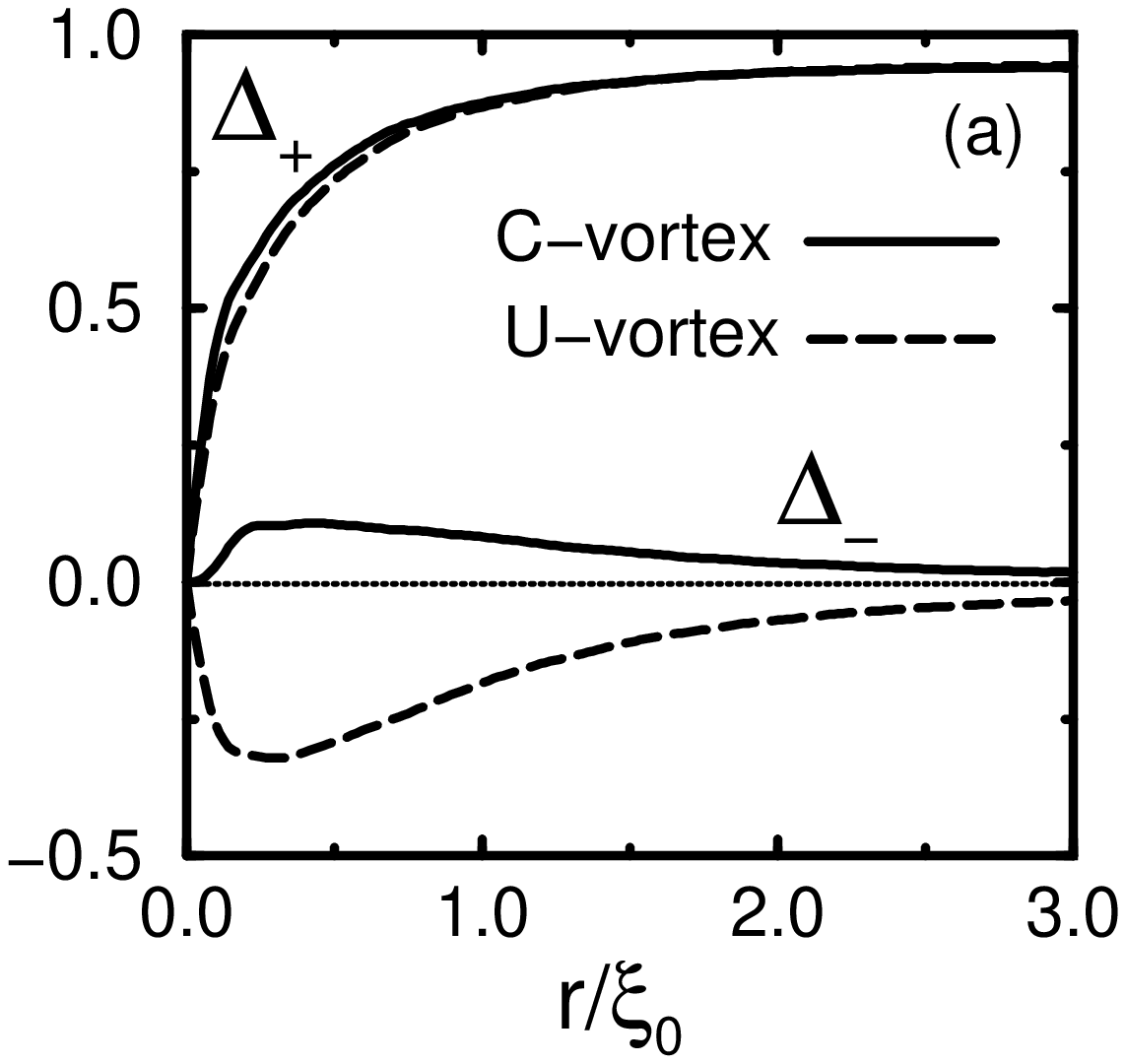}
\end{minipage}
\begin{minipage}{7cm}
\epsfxsize=7cm
\epsfbox{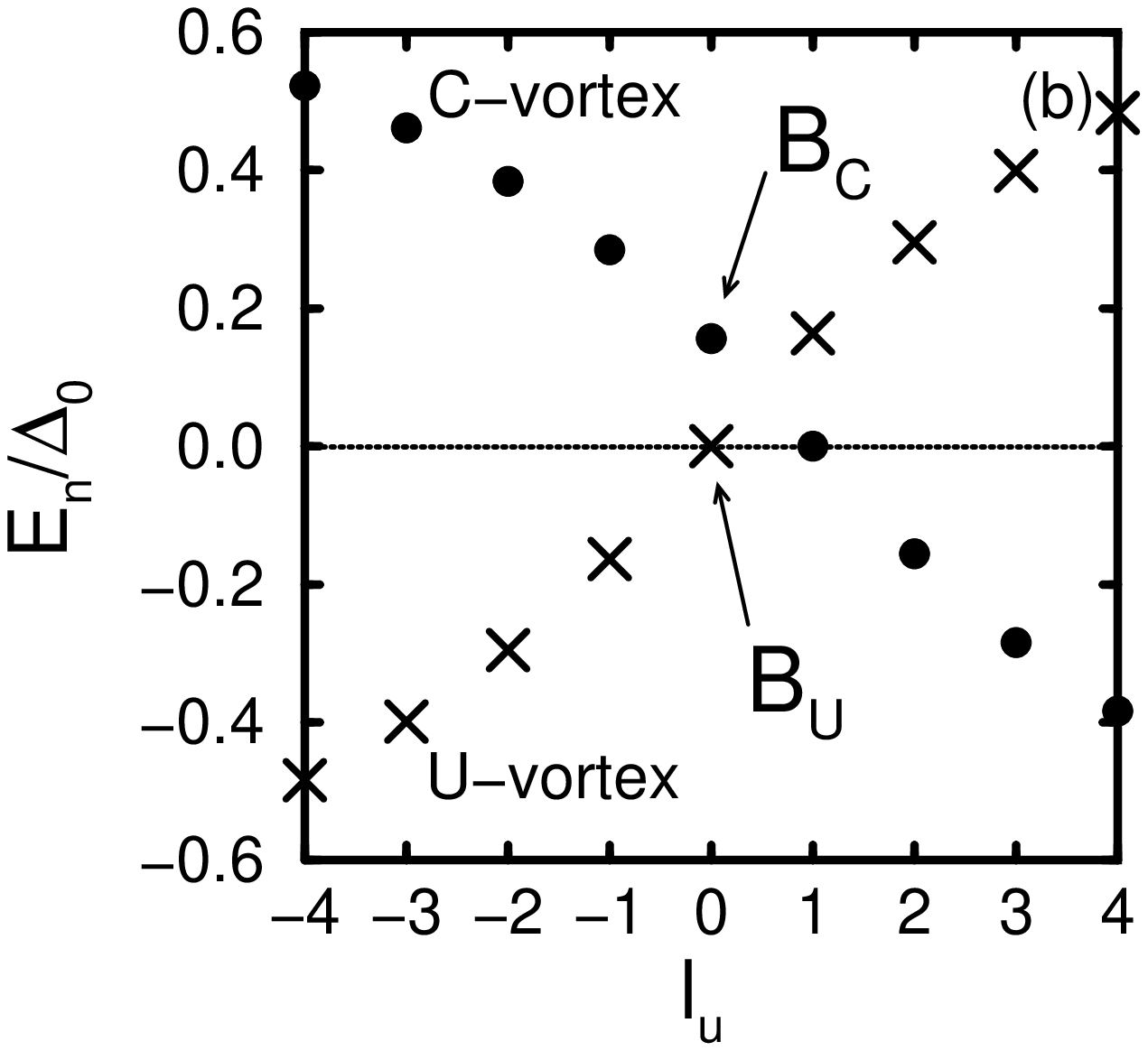}
\end{minipage}
\end{center}
\caption{
Self-consistent results at $T=0$.
(a) Order parameters scaled by $\Delta_0$,
which is the order parameter at $T=0$ in the bulk region.
$\xi_0=v_F/\Delta_0$ is the coherence length.
Set of parameters are taken as $R=8\xi_0$, $k_F\xi_0=16$, $\lambda_{\rm L}=\xi_0$,
$\lambda_{\rm TF}=1/k_{\rm F}$, $E_{\rm c}=\mu$.
Here $\lambda_{\rm L}$ and $\lambda_{\rm TF}$ are the London and Thomas-Fermi screening lengths,
respectively.
(b) Bound state energy spectrum for the C-vortex (circle) and U-vortex (X).
$l_u$ is the angular momentum of the wave function $u_n$.
The bound states represented by ${\rm B_C}$ and ${\rm B_U}$
are important in the vortex charging effect (see the text).
Extended states lie continuously in a $|E_n|\ge \Delta_0$ region.
\protect\cite{Matsumoto99}
}
\label{fig:1}
\end{figure}

\begin{figure}[t]
\begin{center}
\begin{minipage}{7cm}
\epsfxsize=7cm
\epsfbox{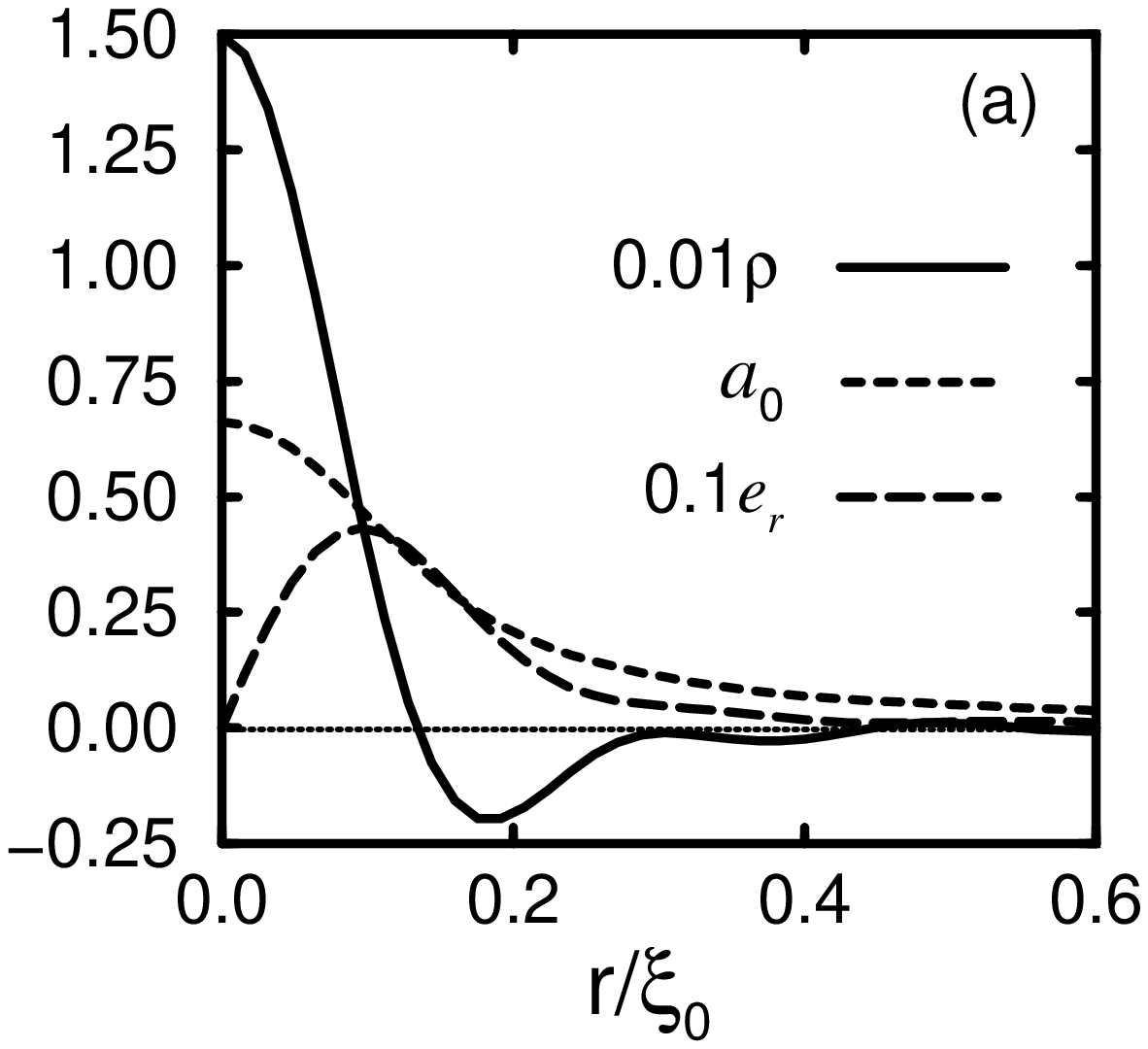}
\end{minipage}
\begin{minipage}{7cm}
\epsfxsize=7cm
\epsfbox{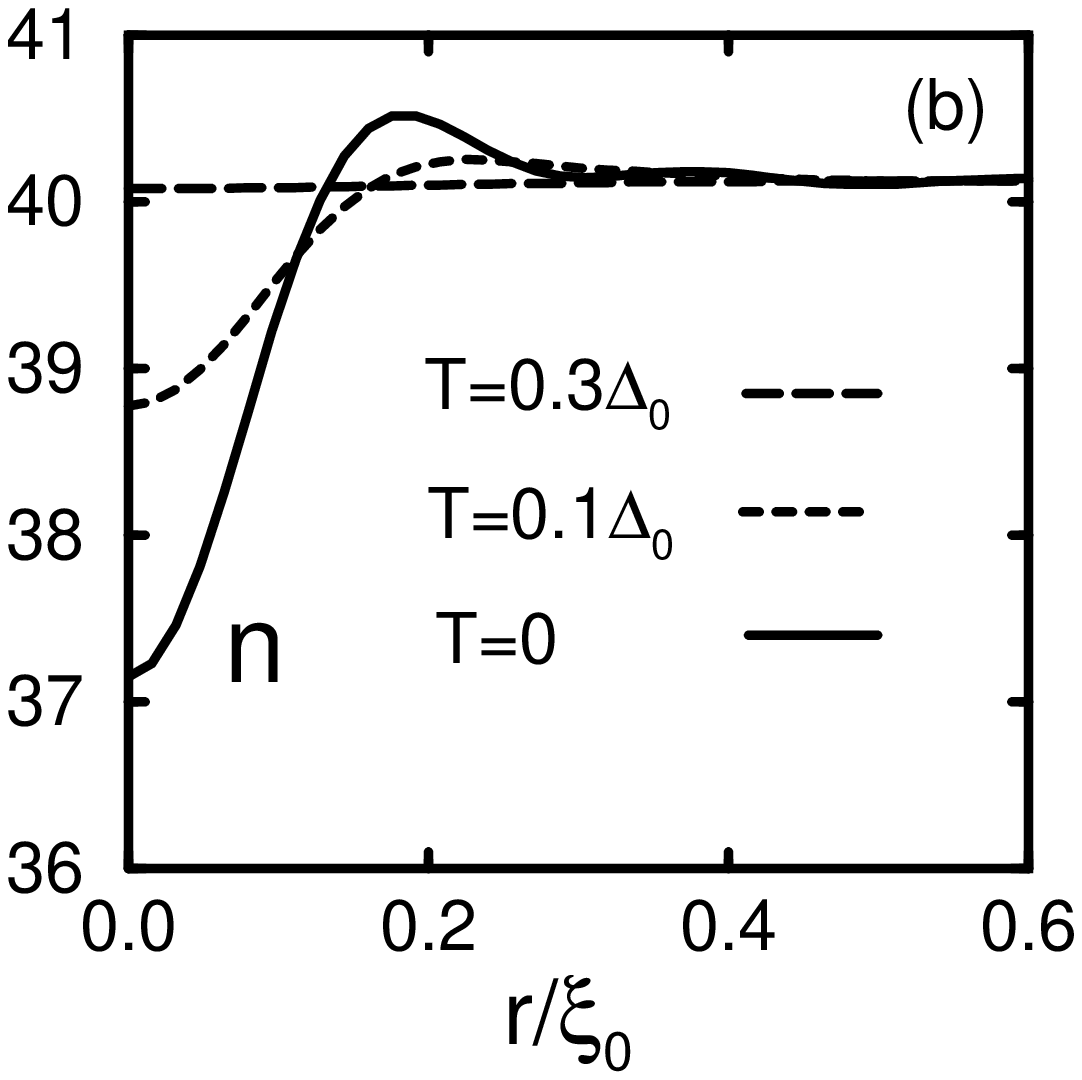}
\end{minipage}
\end{center}
\caption{
(a) Dimensionless scalar potential $a_0=e A_0/\Delta_0$,
electric field $e_r=-\xi_0\partial_r a_0$,
and charge density $\rho=\xi_0(1/r+\partial_r)e_r$ for the C-vortex at $T=0$.
(b) Spatial dependence of the electron density $n$ in $1/\xi_0^2$ unit at various temperatures.
Charge density is given by $\rho(r)=e[n(\infty)-n(r)]$.
}
\label{fig:2}
\end{figure}

Let us discuss our BdG self-consistent solutions.
First we show the order parameters in Fig. \ref{fig:1}(a).
The admixed component $\Delta_-$ is induced around the vortex core as expected.
It shows the asymptotic behavior $\Delta_- \propto r$ ($\Delta_- \propto r^3$ )
for the U-vortex (C-vortex), which is consistent with the GL result.
\cite{Heeb99}
$|\Delta_-|$ for the U-vortex is larger than the C-vortex one.
This indicates that the U-vortex gains much condensation energy.
In Fig. \ref{fig:1}(b) we show several energy eigenvalues of the BdG equation.
We notice that both vortices have zero energy bound states.
\cite{Volovik99}
The appearance of the zero energy states is the consequence
of the symmetric property of the BdG equation [Eq. (\ref{eqn:symmetry})].
As we discuss later,
the bound state for $l_u=0$ is very important to the vortex charging effect,
\cite{Hayashi98-2,Matsumoto00}
where $l_u$ is the angular momentum of the wavefunction $u_n$.
${\rm B_C}$ and ${\rm B_U}$ in Fig. \ref{fig:1}(b) represent the bound state.
Next we show the charge density around the vortex core.
As in Fig. \ref{fig:2}(a) large charge density appears in the vortex core.
\cite{Matsumoto00}
The induced electric field is screened as we go far from the vortex center.
Figure \ref{fig:2}(b) is the spatial dependence of the electron density at various temperatures.
At $T=0$ the electron density is suddenly decreased in the core region,
which results in the vortex charge.
With the increase of temperature, the electron density becomes uniform
and the charge density is reduced accordingly.
Contrary to the C-vortex,
the electron density is almost uniform at all temperatures for the U-vortex,
so that the vortex charge is very small in this case.

\begin{figure}[t]
\begin{center}
\begin{minipage}{7cm}
\epsfxsize=7cm
\epsfbox{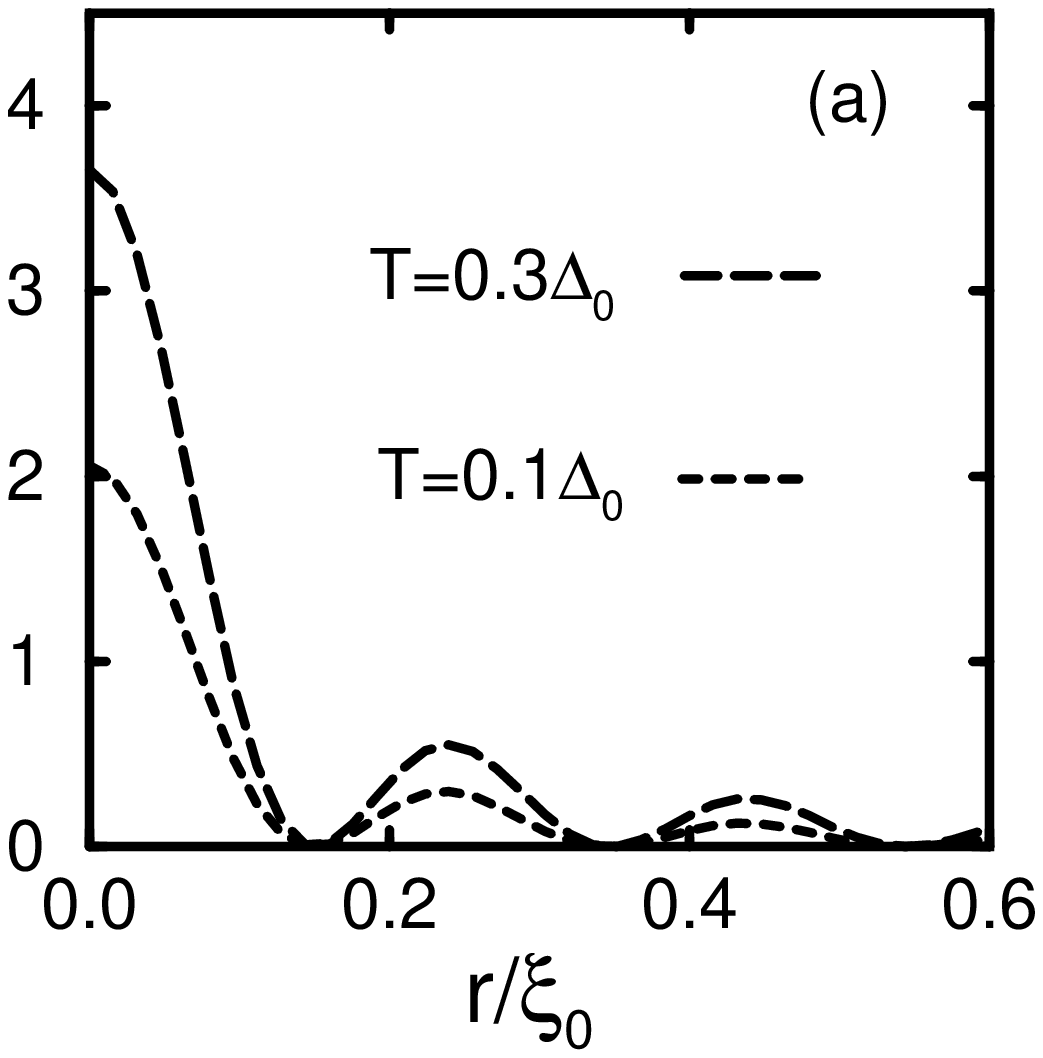}
\end{minipage}
\begin{minipage}{7cm}
\epsfxsize=7cm
\epsfbox{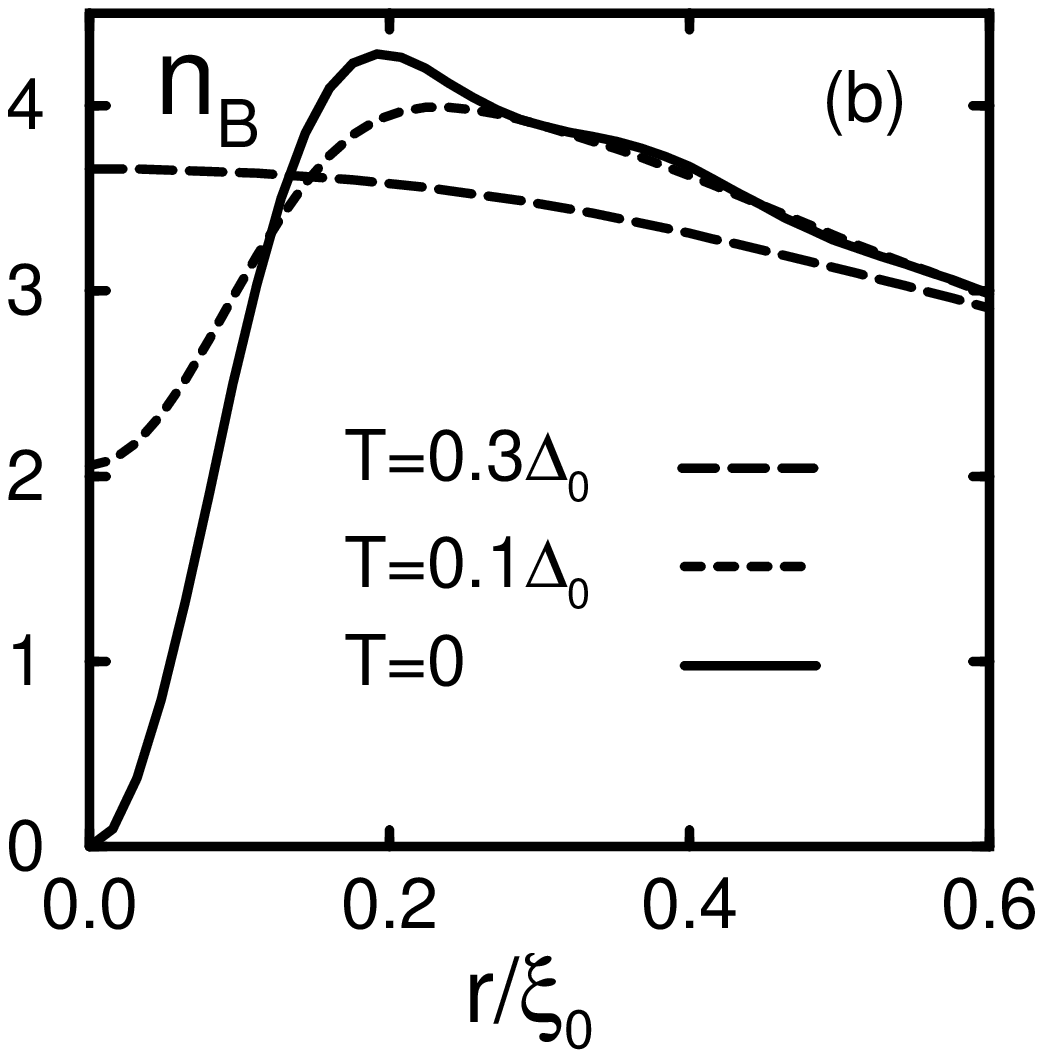}
\end{minipage}
\end{center}
\caption{
Local electron density for the C-vortex in $1/\xi_0^2$ unit.
(a) Contribution from the bound state ${\rm B_C}$ [Fig. \ref{fig:1}(b)].
It is zero at $T=0$.
(b) Bound state contribution $n_{\rm B}$.
}
\label{fig:3}
\end{figure}

Let us explain the microscopic origin of the vortex charging effect.
First we discuss the C-vortex case.
There are two contributions to the electron density $n$.
One is from the bound states, and the other is from the extended states.
We name the former $n_{\rm B}$, and the latter $n_{\rm E}$ ($n=n_{\rm B}+n_{\rm E}$).
In general only the $l_u=0$ states contribute to the local electron density at the vortex center.
In the microscopic study we find out the important role of the bound state
with zero angular momentum ($l_u=0$).
\cite{Hayashi98-2,Matsumoto00}
At zero temperature only the $E_n\le 0$ states are effective in Eq. (\ref{eqn:density}).
Therefore the bound state ${\rm B_C}$ in Fig. \ref{fig:1}(b)
cannot contribute to the electron density, since this state is unoccupied [Fig. \ref{fig:3}(a)].
$n_{\rm B}$ is then suddenly decreased in the core region as in Fig. \ref{fig:3}(b),
and the total electron density is decreased close to the vortex center,
resulting in the finite vortex charge.
At finite temperatures the contribution from the ${\rm B_C}$ state comes out
[Fig. \ref{fig:3}(a)] due to the finite Fermi distribution function
in Eq. (\ref{eqn:density}) for $E_n > 0$.
Correspondingly, $n_{\rm B}$ increases in the core region as shown in Fig. \ref{fig:3}(b).
At $T=0.3\Delta_0$, which is larger than the energy of ${\rm B_C}$,
the electron density is almost uniform due to the contribution from the ${\rm B_C}$ state.
The vortex charge is then reduced strongly at high temperatures as shown in Fig. \ref{fig:2}(b).

Next we discuss the U-vortex.
Contrary to the C-vortex,
the ${\rm B_U}$ bound state in Fig. \ref{fig:1}(b)
can contribute to the electron density at all temperatures,
since this state is located at zero energy.
As the consequence, the electron density for the U-vortex is almost uniform.
In the U-vortex case we find very small temperature dependence in the electron density,
since the ${\rm B_U}$ state is located at zero energy and participates in the energy excitations
with the same rate at all temperatures.
We note that the electron density for the U-vortex is very similar
to that for the C-vortex at $T=0.3\Delta_0$, where the ${\rm B_C}$ state is occupied.
We conclude that the appearance of the vortex charge
depends on the position of the $l_u=0$ bound state relative to the temperature.
In the conventional $s$-wave case,
the vortex charge always appears at sufficiently low temperatures,
\cite{Hayashi98-2}
which is similar to the C-vortex result.

\begin{figure}[t]
\begin{center}
\begin{minipage}{7cm}
\epsfxsize=7cm
\epsfbox{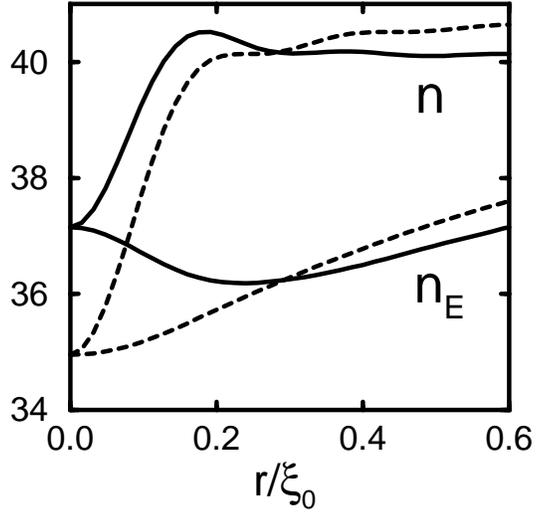}
\end{minipage}
\end{center}
\caption{
Total electron density $n$ and contribution from the extended states $n_{\rm E}$ at $T=0$.
Solid lines represent the result with charge screening ($\lambda_{\rm TF}=1/k_{\rm F}$),
while dashed lines are the result without charge screening ($\lambda_{\rm TF}\rightarrow \infty$).
}
\label{fig:4}
\end{figure}

Let us mention the effect of charge screening.
We show the total electron density $n$ and contribution from extended states $n_{\rm E}$
for two cases, where charge screening is taken into account or not (see Fig. \ref{fig:4}).
In the no charge screening case,
the contribution from extended states $n_{\rm E}$ is small at the vortex core.
The total electron density $n$ is then decreased, resulting in larger vortex charge.
Note that the scalar potential in Eq. (\ref{eqn:BdG2}g) is zero ($\bar{a}_0=0$)
for no charge screening ($\lambda_{\rm T} \rightarrow \infty$).
Therefore the scalar potential does not work to screen the vortex charge.
On the other hand, when the charge screening turns on,
quasiparticles respond to the vortex charge via the scalar potential $\bar{a}_0$
in the BdG equation (\ref{eqn:BdG2}a)-(\ref{eqn:BdG2}c).
$n_{\rm E}$ is then increased to cancel the vortex charge.
In contract to the extended state,
electron density of the bound states $n_{\rm B}$ does not exhibit
difference between the charge screening and no screening cases.
Since the bound states are localized around the vortex core,
their wave function hardly modulate, while the extended states do easily.
Hence the extended states quickly respond to the electric field and they screen the vortex charge.
However, there still remains the substantial vortex charge.

\begin{figure}[t]
\begin{center}
\begin{minipage}{12cm}
\epsfxsize=12cm
\epsfbox{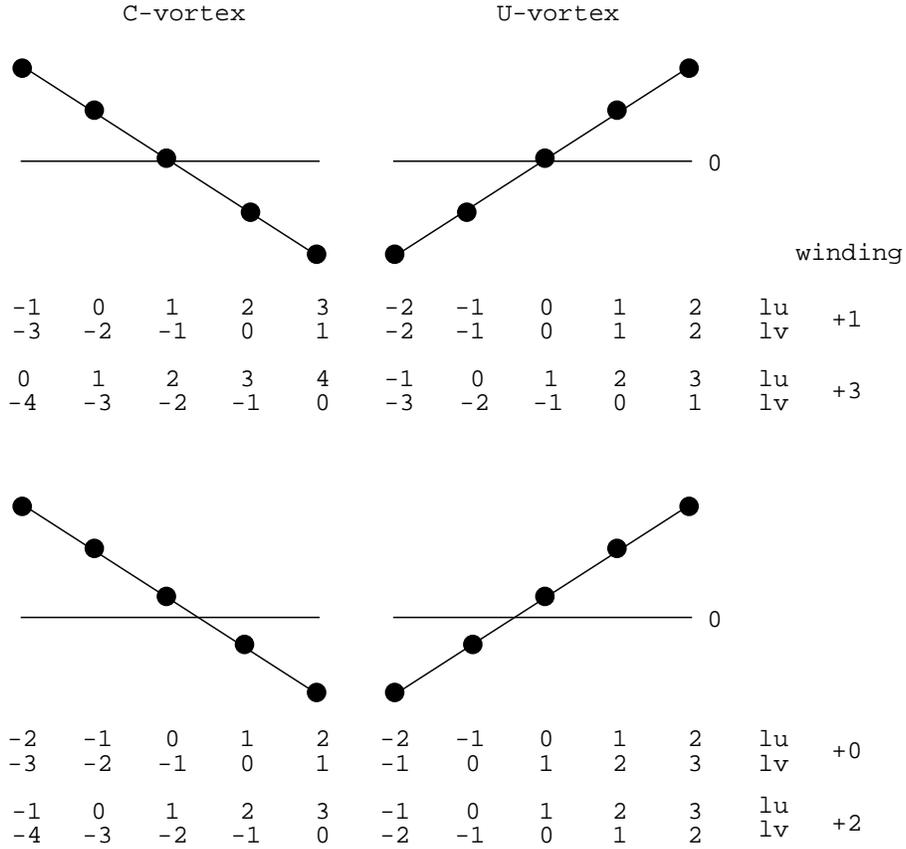}
\end{minipage}
\end{center}
\caption{
Schematic picture of bound state energy spectrum for various winding pairings.
Numbers represent the angular momenta of $u_n$ and $v_n$ ($l_u$ and $l_v$).
}
\label{fig:5}
\end{figure}

We discuss the effect of magnetic field.
The Chern-Simons physics
\cite{Goryo99}
and Bernoulli effect
\cite{Ivanchenko}
take place in the presence of magnetic field.
To see this effect we compare the two results,
where $\bar{\bbox{a}}$ is present or ignored (no magnetic field).
The scalar potential $\bar{a}_0$ is taken into account for both cases.
Our result shows that the effect of the magnetic field is too small to be seen.
Thus we find that the vortex charge is mainly determined
by the microscopic quasiparticles structure,
which reflects the phase winding of the chiral pairing.
Figure \ref{fig:5} shows a schematic picture of the energy spectrum of vortex bound states
for various winding pairings.
The lowest-lying state appears at the symmetric position of $l_u$ and $l_v$
due to the relation (\ref{eqn:symmetry}),
where $l_u$ and $l_v$ are the angular momenta of $u_n$ and $v_n$, respectively.
For the C-vortex the bound state energy is positive for $l_u=0$,
while it is negative or zero for the U-vortex.
Therefore the C-vortex is always charged up at low temperatures,
while the U-vortex is not.
Thus the vortex charge can be discussed in terms of the vortex bound states
for all types of chiral pairings.
Similarly, we can understand that the $s$-wave (zero winding) vortex is charged up.
\cite{Hayashi98-2}

Finally let us discuss non-chiral pairings, such as $p_x$, $p_y$, and $d_{x^2-y^2}$-waves.
These states are all gapless
and it is reported that there is no bound states for the $d_{x^2-y^2}$-wave.
\cite{Franz}
Since the system is anisotropic in this case,
the energy eigenstates are not classified by the angular momentum.
Therefore the above argument for the vortex charging is difficult to apply to this case.
However, our result implies that the detailed quasiparticle structure in the vortex core
is important for the vortex charging effect even in the gapless pairing cases.

\section{Summary and Discussions}
In conclusion, we have solved the problem of a single $p_x\pm i p_y$-wave vortex
self-consistently within the Bogoliubov-de Gennes theory.
The full self-consistent calculation including both charge and current screenings
was performed for the first time to investigate vortex problems.
Due to the time reversal symmetry breaking, there are two types of vortices.
We found a substantial vortex charge in the C-vortex case,
while the vortex charge is suppressed for U-vortex.
We conclude that the vortex charging effects are mainly determined
by the local quasiparticle structure around the vortex core,
reflecting the chiral $p_x\pm i p_y$-wave paring.
Especially the lowest vortex bound state is very important for the charging effect
at low temperatures.

In a real sample we can expect two types ($p_x \pm i p_y$) of domains.
Therefore we expect both C and U-vortices in the presence of an external magnetic field.
In domains, where the C-vortex is realized,
an electronic field is induced due to the vortex charge.
Therefore the charge of the C-vortex can be detected.
If we use a field cooled sample, a single domain forming the U-vortex is realized
all over the sample,
since the U-vortex is energetically favorable.
\cite{Heeb99}
In this case it is difficult to find a signal from the U-vortex,
since the charge of the U-vortex is much smaller than the C-vortex one.
Thus we can distinguish the two types of C and U-vortices.

How can we detect the vortex charge?
The vortex charge can induce lattice distortion and it scatters neutrons.
It is reported that a polarized neutron scattering can be used to detect the vortex charge.
\cite{Neumann}
NQR is also one of the possible experiment which can detect the vortex charging effect,
since the NQR detects the local electric field induced by the vortex charge.
Very recently, Kumagai $et$ $al$. reported that the vortex charge is observed by the NQR
in a high-$T_c$ material.
\cite{Kumagai}
Thus the detection of the vortex charge is in progress now.
It is very exciting if the vortex charge of the chiral superconductor is detected,
since it can provide a very strong evidence of the {\it rotating} Cooper pair.

\acknowledgements
The authors express their sincere thanks to M. Sigrist, G. Blatter, and A. Furusaki
for many stimulating discussions.
One of the authors (MM) thanks N. Hayashi for helpful discussions
on the BdG equation and the vortex charging effect.
He would like to acknowledge M. Koga for his critical reading of the manuscript.
This work is supported by Casio Science Promotion Foundation.

\appendix
\section{Derivation of Bogoliubov-de Gennes equation and Gap equation}
In this appendix we derive the Bogoliubov-de Gennes equation
and gap equation for the $p_x + i p_y$-wave superconductor, which we use in the text.

Let us start from the the following Bogoliubov-de Gennes equation:
\begin{mathletters}
\begin{eqnarray}
 h_0(\bbox{r}) u_n(\bbox{r})+\int d \bbox{r}'\Delta(\bbox{r},\bbox{r}')v_n(\bbox{r}')&=&E_n u_n(\bbox{r}), \\
-h_0^*(\bbox{r}) v_n(\bbox{r})-\int d \bbox{r}'\Delta^*(\bbox{r},\bbox{r}')u_n(\bbox{r}')&=&E_n v_n(\bbox{r}).
\end{eqnarray}
\label{eqn:A.1}
\end{mathletters}\noindent
Coordinates $\bbox{r}$ and $\bbox{r}'$ can be transformed
into the center of mass and relative coordinates.
\begin{equation}
\bbox{R}={\bbox{r}+\bbox{r}' \over 2},~~~~~~\bbox{X}=\bbox{r}-\bbox{r}'.
\end{equation}
The $p_x +i p_y$-wave pairing is expressed as
\begin{mathletters}
\begin{eqnarray}
\Delta(\bbox{R},\bbox{k}) &=& \Delta_x(\bbox{R}){k_x \over k_{\rm F}}
                          +i\Delta_y(\bbox{R}){k_y \over k_{\rm F}}, \\
\Delta(\bbox{R},\bbox{X}) &=& {1 \over (2\pi)^2}\int d \bbox{k}
          e^{i\mbox{\footnotesize \boldmath $k$}\cdot\mbox{\footnotesize \boldmath $X$}}
          \Delta(\bbox{R},\bbox{k}) \cr
&=&i{1 \over (2\pi)^2}\int d \bbox{k} {1 \over k_{\rm F}}
                            \bigl[\Delta_x(\bbox{R}) \partial_{x'}
 +i\Delta_y(\bbox{R}) \partial_{y'}\bigr]
   e^{i\mbox{\footnotesize \boldmath $k$}\cdot(\mbox{\footnotesize \boldmath$r$}
   -\mbox{\footnotesize \boldmath$r$}')} \cr
&=& {i \over k_{\rm F}}\bigl[\Delta_x(\bbox{R})\partial_{x'}+i\Delta_y(\bbox{R})\partial_{y'}\bigr]
                  \delta(\bbox{r}-\bbox{r}').
\end{eqnarray}
\label{eqn:A.3}
\end{mathletters}\noindent
Here $\Delta_x$ and $\Delta_y$ are the $p_x$ and $p_y$ components of the order parameter.
Substituting Eq. (\ref{eqn:A.3}b) into Eq. (\ref{eqn:A.1}), we obtain
\begin{mathletters}
\begin{eqnarray}
&&h_0(\bbox{r}) u_n(\bbox{r})
  -{{\rm i} \over k_{\rm F}}\Bigl\{
    \Delta_x(\bbox{r}) \partial_x +i\Delta_y(\bbox{r}) \partial_y
    +{1 \over 2}\bigl[ \partial_x \Delta_x(\bbox{r})+i \partial_y \Delta_y(\bbox{r})\bigr]
  \Bigr\}v_n(\bbox{r})=E_n u_n(\bbox{r}), \\
-&&{h_0}^*(\bbox{r}) v_n(\bbox{r})
  -{i \over k_{\rm F}}\Bigl\{
    \Delta_x(\bbox{r}) \partial_x -i\Delta_y(\bbox{r}) \partial_y
    +{1 \over 2}\bigl[ \partial_x \Delta_x(\bbox{r})-i \partial_y \Delta_y(\bbox{r})\bigr]
\Bigr\}u_n(\bbox{r})=E_n v_n(\bbox{r}).
\end{eqnarray}
\end{mathletters}\noindent

Next we derive gap equation.
The order parameter is defined by
\begin{equation}
\Delta(\bbox{r},\bbox{r}')=V(\bbox{r},\bbox{r}')
    \sum_{0\le E_n \le E_{\rm c}} \tanh\left(\frac{E_n}{2T}\right)
    u_n(\bbox{r}'){v_n}^*(\bbox{r})\equiv V(\bbox{r},\bbox{r}')D(\bbox{r},\bbox{r}'),
\end{equation}
where $E_{\rm c}$ is a cutoff energy.
For the $p$-wave pairing, we assume the following $p$-wave interaction:
\begin{equation}
V(\bbox{r},\bbox{r}')=-{V_p \Omega \over (2\pi)^2}\int d \bbox{k}
    e^{i\mbox{\footnotesize \boldmath $k$}\cdot\mbox{\footnotesize \boldmath $X$}}
       {k^2 \over k_{\rm F}^2}.
\end{equation}
Here $V_p>0$ and $\Omega$ are the $p$-wave attractive interaction
and two-dimensional volume of the system, respectively.
The order parameter is then expressed as
\begin{mathletters}
\begin{eqnarray}
\Delta(\bbox{R},\bbox{k})&=&\int d \bbox{X} e^{-i\mbox{\footnotesize \boldmath $k$}\cdot\mbox{\footnotesize \boldmath $X$}}\Delta(\bbox{r},\bbox{r}') \cr
&=&-V_p \Omega \int d \bbox{k}'{k'^2 \over k_{\rm F}^2} {1 \over (2\pi)^2}\int d \bbox{X}
e^{-i\mbox{\footnotesize \boldmath $k$}\cdot\mbox{\footnotesize \boldmath $X$}}
D(\bbox{R}+{\bbox{X} \over 2},\bbox{R}-{\bbox{X} \over 2}) e^{i\mbox{\footnotesize \boldmath $k$}'\cdot\mbox{\footnotesize \boldmath $X$}}.
\end{eqnarray}
\end{mathletters}\noindent
We expand $D(\bbox{R}+{\bbox{X} \over 2},\bbox{R}-{\bbox{X} \over 2})$ as
\begin{equation}
D(\bbox{R}+{\bbox{X} \over 2},\bbox{R}-{\bbox{X} \over 2})\simeq R(\bbox{R},\bbox{R})
+\bigl[{\partial D(\bbox{R},\bbox{R}') \over \partial \bbox{R}}
      -{\partial D(\bbox{R},\bbox{R}') \over \partial \bbox{R}'}
      \bigr]_{\mbox{\footnotesize \boldmath $R$}'\rightarrow\mbox{\footnotesize \boldmath $R$}}
      \cdot{\bbox{X} \over 2},
\end{equation}
and obtain
\begin{eqnarray}
\Delta(\bbox{R},\bbox{k}) &\simeq& - V_p \Omega {k^2 \over k_{\rm F}^2}
           \sum_{0 \le E_n \le E_{\rm c}} \tanh\left(\frac{E_n}{2T}\right)
           u_n(\bbox{R})v_n^*(\bbox{R}) \cr
       &-& i V_p \Omega {\bbox{k} \over k_{\rm F}^2}\cdot({\partial \over \partial \bbox{R}}
          -{\partial \over \partial \bbox{R}'})
      \sum_{0 \le E_n \le E_{\rm c}} \tanh\left(\frac{E_n}{2T}\right) u_n(\bbox{R})v_n^*(\bbox{R}')
          |_{\mbox{\footnotesize \boldmath $R$}'\rightarrow\mbox{\footnotesize \boldmath $R$}}.
\label{eqn:gap}
\end{eqnarray}
The first term in Eq. (\ref{eqn:gap}) is zero due to the $p$-wave pairing.
Since $\Delta(\bbox{R},\bbox{k})$ can be divided into two parts as in Eq. (\ref{eqn:A.3}a),
the gap equation takes the following form:
\begin{mathletters}
\begin{eqnarray}
\Delta_x(\bbox{r}) &=& - V_p \Omega {i \over k_{\rm F}}( \partial_x - \partial_{x'})
    \sum_{0 \le E_n \le E_{\rm c}} \tanh\left(\frac{E_n}{2T}\right)
    u_n(\bbox{r}){v_n}^*(\bbox{r}')|_{\bbox{r}'\rightarrow\bbox{r}}, \\
\Delta_y(\bbox{r}) &=& - V_p \Omega {1 \over k_{\rm F}}( \partial_y - \partial_{y'})
    \sum_{0 \le E_n \le E_{\rm c}} \tanh\left(\frac{E_n}{2T}\right)
    u_n(\bbox{r}){v_n}^*(\bbox{r}')|_{\bbox{r}'\rightarrow\bbox{r}}.
\end{eqnarray}
\end{mathletters}\noindent
For a cylindrical system,
it is convenient to introduce the following form for order parameters:
\begin{equation}
\Delta_\pm(\bbox{r})
=\frac{1}{2}[\Delta_x(\bbox{r})\pm\Delta_y(\bbox{r})].
\end{equation}
Here $\Delta_\pm$ is the order parameter for the $p_x \pm p_y$-wave pairing.
By using the $\Delta_\pm$ representation,
the Bogoliubov-de Gennes equation and the gap equations are expressed as
\begin{mathletters}
\begin{eqnarray}
&&h_0(\bbox{r}) u_n(\bbox{r})
  -{{\rm i} \over k_{\rm F}}\Bigl\{
    \Delta_+(\bbox{r})\Box_+ +\Delta_-(\bbox{r})\Box_-
    +{1 \over 2}\bigl[\Box_+ \Delta_+(\bbox{r})+\Box_- \Delta_-(\bbox{r})\bigr]
    \Bigr\}v_n(\bbox{r})=E_n u_n(\bbox{r}), \\
-&&h_0^*(\bbox{r}) v_n(\bbox{r})-{i \over k_{\rm F}}\Bigl\{
    \Delta_+(\bbox{r})\Box_+ +\Delta_-(\bbox{r})\Box_-
    +{1 \over 2}\bigl[\Box_+ \Delta_+(\bbox{r})+\Box_- \Delta_-(\bbox{r})\bigr]
    \Bigr\}^*u_n(\bbox{r})=E_n v_n(\bbox{r}), \\
&&\Delta_\pm(\bbox{r})=- V_p \Omega {i \over 2k_{\rm F}}(\Box_\mp-\Box'_\mp \bigr)
    \sum_{0 \le E_n \le E_{\rm c}} \tanh\left(\frac{E_n}{2T}\right)
    u_n(\bbox{r}){v_n}^*(\bbox{r}')|_{\bbox{r}'\rightarrow\bbox{r}}, \\
&&\Box_\pm= \partial_x \pm i \partial_y.
\end{eqnarray}
\end{mathletters}\noindent




\vspace{3cm}

\end{document}